\documentclass[preprint,number,sort&compress, numafflabel]{elsarticle}

\usepackage{epsfig}
\usepackage{graphicx}
\usepackage{lineno}
\linenumbers

\usepackage{epstopdf} 

\journal{Nuclear Instruments \& Methods  A}

\begin{document}
\begin{frontmatter}

\title{Improvement in Fast Particle Track Reconstruction with Robust Statistics}

\author[Adelaide]{M.~G.~Aartsen}
\author[MadisonPAC]{R.~Abbasi}
\author[Gent]{Y.~Abdou}
\author[Zeuthen]{M.~Ackermann}
\author[Christchurch]{J.~Adams}
\author[Geneva]{J.~A.~Aguilar}
\author[MadisonPAC]{M.~Ahlers}
\author[Berlin]{D.~Altmann}
\author[MadisonPAC]{J.~Auffenberg}
\author[Bartol]{X.~Bai\fnref{SouthDakota}}
% ...
\address[Adelaide]{School of Chemistry \& Physics, University of Adelaide, Adelaide SA, 5005 Australia}
% ...
\address[Berlin]{Institut f\"ur Physik, Humboldt-Universit\"at zu Berlin, D-12489 Berlin, Germany}
% ...
\address[Christchurch]{Dept.~of Physics and Astronomy, University of Canterbury, Private Bag 4800, Christchurch, New Zealand}
% ...
\address[Geneva]{D\'epartement de physique nucl\'eaire et corpusculaire, Universit\'e de Gen\`eve, CH-1211 Gen\`eve, Switzerland}
\address[Gent]{Dept.~of Physics and Astronomy, University of Gent, B-9000 Gent, Belgium}
% ...
\address[MadisonPAC]{Dept.~of Physics and Wisconsin IceCube Particle Astrophysics Center, University of Wisconsin, Madison, WI 53706, USA}
% ...
\address[Bartol]{Bartol Research Institute and Department of Physics and Astronomy, University of Delaware, Newark, DE 19716, USA}
% ...
\address[Zeuthen]{DESY, D-15735 Zeuthen, Germany}
% ...
\fntext[SouthDakota]{Physics Department, South Dakota School of Mines and Technology, Rapid City, SD 57701, USA}
\fntext[LosAlamos]{Los Alamos National Laboratory, Los Alamos, NM 87545, USA}
\fntext[Bari]{also Sezione INFN, Dipartimento di Fisica, I-70126, Bari, Italy}
\fntext[Goddard]{NASA Goddard Space Flight Center, Greenbelt, MD 20771, USA}
\cortext[cor1]{Corresponding author.  Email: wellons@icecube.wisc.edu, Phone: 304-542-4464, Address: Wisconsin Institutes for Discovery, 
330 N. Orchard St., 
Madison, WI 53715}
 
\begin{abstract}

The IceCube project has transformed one cubic kilometer of deep natural Antarctic ice into a Cherenkov detector.  Muon neutrinos are detected and their direction inferred by mapping the light produced by the secondary muon track inside the volume instrumented with photomultipliers. Reconstructing the muon track from the observed light is challenging due to noise, light scattering in the ice medium, and the possibility of simultaneously having multiple muons inside the detector, resulting from the large flux of cosmic ray muons.

This manuscript describes  work on two problems: (1) the \emph{track
  reconstruction} problem, in which, given a set of observations, the
goal is to recover the track of a muon; and (2) the
\emph{coincident event} problem, which is to determine how many
muons are active in the detector during a time window.   Rather than solving these problems by developing more complex physical models that are applied at later stages of the analysis, our approach is to augment the detector's early reconstruction with data filters and robust statistical techniques. These can be implemented at the level of on-line reconstruction and, therefore, improve all subsequent reconstructions.   Using the metric of median angular resolution,  a
standard metric for track reconstruction, we improve the
accuracy in the initial reconstruction direction by 13\%.  We also present improvements in measuring the number of muons in coincident events: we can accurately determine the number of muons 98\% of the time. 

\end{abstract}
\end{frontmatter}

\section{Introduction}\label{sec:intro}
\begin{figure}[tbp]
\begin{center}
\epsfig{file=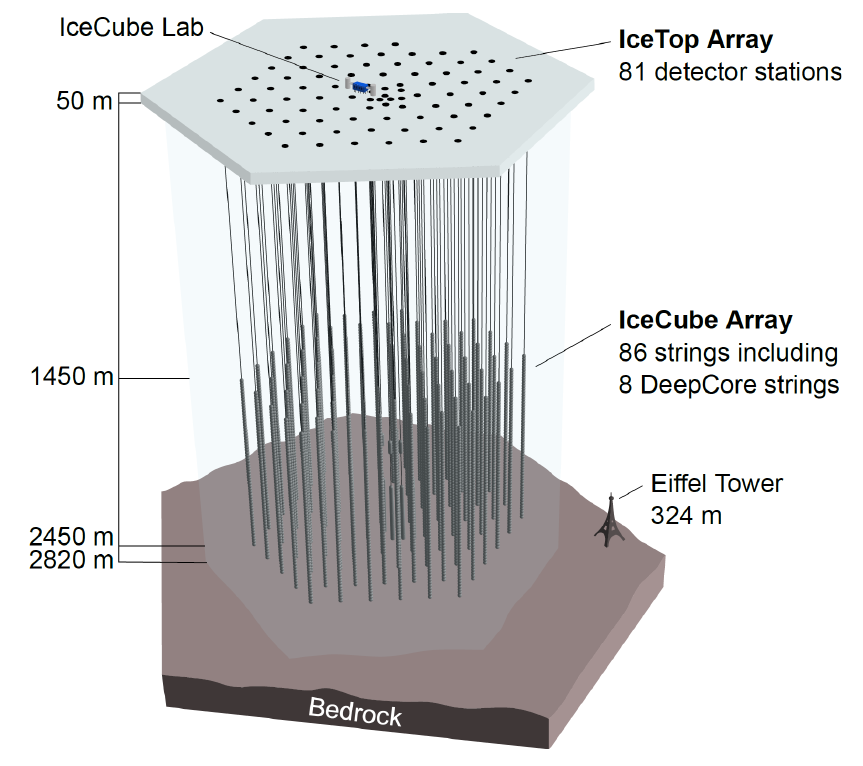,height = 7 cm}
\caption{The IceCube neutrino detector in the Antarctic ice.  A picture of the Eiffel Tower is shown for scale.}
\label{fig:IceCube detector}
\end{center}
\end{figure}

The IceCube neutrino detector searches for neutrinos that are generated by the universe's most
violent astrophysical events: exploding stars, gamma ray bursts, and
cataclysmic phenomena involving black holes and neutron stars~\cite{icecubePurpose}. The detector, roughly one cubic kilometer in size,  is located near the geographic South Pole and is buried to a depth of about
2.5~km in the Antarctic ice~\cite{DetectorGeneralInfo}.  The
detector is illustrated in Figure \ref{fig:IceCube detector}, and a
more complete description is given in Section~\ref{sec:background}.

When a neutrino enters the telescope, it occasionally interacts in the ice and generates a muon.  The neutrino direction can be inferred from a reconstruction of the muon track.  Muons are also generated by cosmic rays interacting in the atmosphere, and separation of the background of cosmic ray muons and neutrino-induced muons is a necessary step for neutrino analysis.  This separation is challenging, as  the number of observed cosmic ray muons exceeds the number of observed neutrino muons by over five orders of magnitude~\cite{detectorBackground}. 

The primary mechanism for separating the cosmic ray muons from the neutrino muons  is reconstructing the muon track and determining whether the muon was traveling downwards into the Earth or upwards out of the Earth.  Because neutrinos can penetrate through the Earth but cosmic ray muons cannot, it follows that a muon traveling out of the Earth must have been generated by a neutrino.  Thus, by selecting only the muons that are reconstructed as up-going, the cosmic ray muons can, in principle, be removed from the data.  Because the number of cosmic ray muons overwhelms the number of neutrino muons,  high accuracy is critical for preventing erroneous reconstruction of cosmic ray muons as neutrino-induced.

Here, we examine two problems that arise in the  separation of cosmic ray muons from neutrino muons in the IceCube detector:\begin{enumerate}

\item {\em Reconstruction}, in which the track of a muon is reconstructed from the observed light at different positions and times in the
  detector.  

\item 
  {\em Coincident Event Detection}, in which we detect the number of muons inside the detector, and assign observed photons to a muon.  
\end{enumerate}

 Sophisticated reconstruction techniques have been developed that computationally model in detail the muon's Cherenkov cone, as well as  the scattering and absorption of photons through layers of Antarctic ice with varying optical properties~\cite{detectorBackground, icelayersNew, opticalProp}. Rather than further refining these techniques, the current work focusses on improving the statistical techniques and optimizing data filtering in the early online track reconstruction performed on the data in real time at the South Pole. Besides benefiting directly any analysis that uses the online reconstruction, such as the search for cosmogenic neutrinos, any later analysis will benefit from improvements made at the early stages of the data collection.

\subsection
{Related Work}
Track reconstruction and coincident event detection challenges are ubiquitous in particle physics~\cite{ ATLAS, multijets, jets}, both in particle accelerators and cosmic particle detectors.  While the work described in this manuscript builds on the previous technique developed for the IceCube detector~\cite{detectorBackground}, these techniques are general purpose and potentially have applications in detectors beyond IceCube.

\subsection
{Outline} 
We begin by describing the IceCube detector and track reconstruction challenges in Section~\ref{sec:background}.  In Section~\ref{sec:pipeline}, we describe the reconstruction pipeline including the prior IceCube software, then we present improvements to the online tracking algorithm and discuss the results.  Section~\ref{sec:coincident} describes improvements on coincident event detection and follows a parallel structure to Section~\ref{sec:pipeline}.  We conclude in Section \ref{sec:conclusions}.

\section{IceCube Detector and Track Reconstruction Challenges}\label{sec:background}

The IceCube detector is composed of 5,160 optical detectors, each containing a photomultiplier tube (PMT) and an onboard digitizer~\cite{PMTs}.  The PMTs are spread over 86 vertical strings arranged in a hexagonal shape, with a total instrumented volume of approximately one cubic kilometer.   The PMTs on a given string are separated vertically by 17~m, and the string-to-string separation is roughly 125~m. 

 At an abstract level, the IceCube detector operates by detecting muons as they travel through the instrumented volume of ice.  As the muon travels through the detector, it radiates light~\cite{icelayersNew}, which is observed by the PMTs and quantized into discrete \emph{hits}~\cite{dataAquiring}. The detector uses several trigger criteria. The most commonly used trigger selects time intervals where eight PMTs (with local coincidences) are fired within 5 microseconds. When a trigger occurs, all data within a 10-microsecond trigger window is saved, becoming an \emph{event}. If the number of hits in an event is sufficiently large, the muon track reconstruction algorithm is triggered.
 
There are several challenges for the  reconstruction algorithms used in the detector.  Varying optical properties of the ice affect reconstruction accuracy, the data may contain outlier hits due to uncorrelated noise, and there are finite computational resources available to tracking code run on-site.

\paragraph{Modeling Difficulties}
 The details of the ice's optical properties are nontrivial to model. Light propagating through the ice is affected by scattering and absorption. These effects cannot be analytically calculated, and the optical properties of the ice vary with depth~\cite{opticalProp}.  In addition, 
      the Cherenkov light originates 
      both directly from the muon and from particles showers initiated  
      by stochastic energy losses of the muon.
\paragraph{Noise} The noise inherent in the data is another challenge.  
Noise hits can arise, either from the thermal background of the photocathode or from photons generated by radioactive decay inside the PMT~\cite{PMTs}. 

\paragraph{Computational Constraints}The reconstruction algorithms are also limited in complexity by the computing resources available at the South Pole.  The track reconstruction algorithm has to process about 3,000 muons per second,  so algorithms with excessive computational demands are discouraged.    


\section{Reconstruction Improvement}\label{sec:pipeline}

As shown in the following, augmenting the reconstruction algorithm with some basic filters and classical data analysis techniques results in significant improvement in the reconstruction algorithm's accuracy.   

\subsection{Prior IceCube Software}

\begin{figure*}[t]
\centering
\epsfig{file=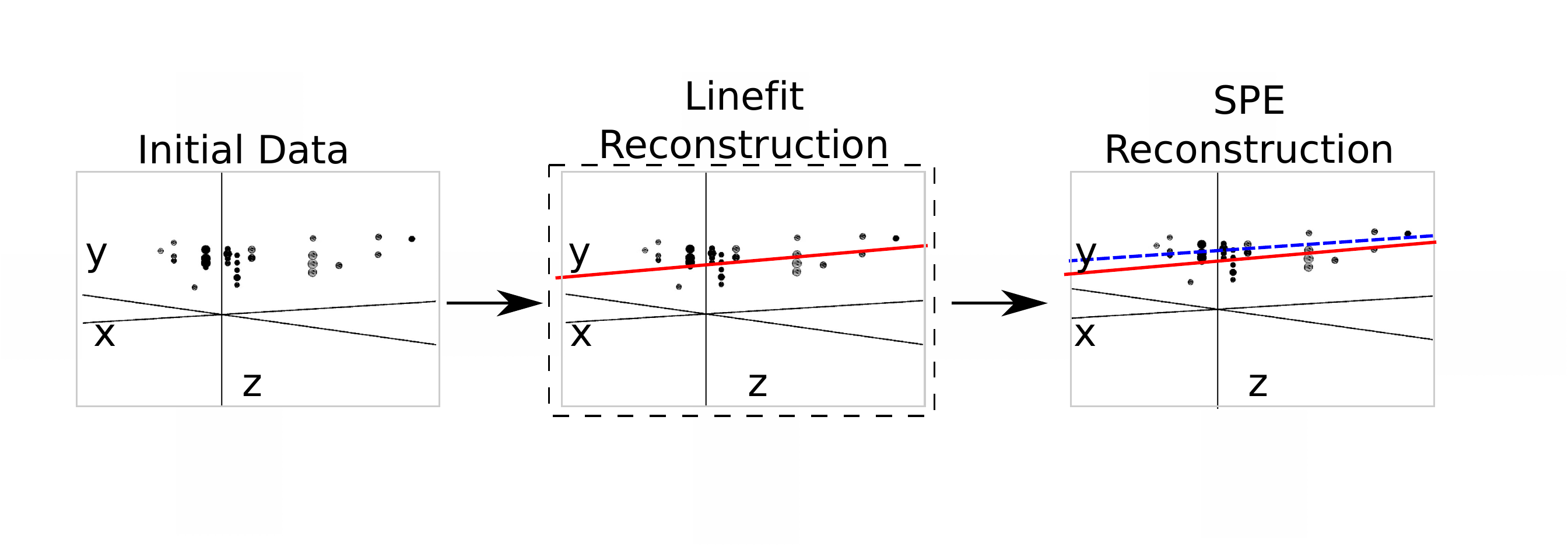, width=13.3 cm}
\caption{ The reconstruction pipeline used to process data in the IceCube detector.  After initial data are collected, it is then processed by some basic noise filters, which remove clear outliers.  This cleaned data are  processed by a basic reconstruction algorithm (solid line), which is used as the seed for the more sophisticated reconstruction algorithm (dashed  line).   The sophisticated reconstruction is then evaluated as a potential neutrino.   \label{fig:simple reconstruction pipeline} The work presented in this manuscript makes changes to the basic reconstruction step (indicated by the dashed   box).   
  \label{fig:improved reconstruction pipeline} }
\end{figure*}

The muon track reconstruction process (outlined in Figure \ref{fig:simple reconstruction pipeline}) starts when the number of detected hits exceeds a preset threshold and initiates data collection.  After the initial data are collected, the event then passes through a series of basic filters to remove obvious outliers~\cite{hitcleaning}.  

This is followed by a basic reconstruction algorithm, \textit{linefit}~\cite{linefitName},   that disregards the Cherenkov cone and, instead, finds the track that minimizes the sum of the squares of the distances between the track and the hits.  More formally, assume there are  $N$ hits;  denote the position and time of the $i$th hit as $\vec{x}_i$ and $t_i$, respectively.    Let the reconstructed muon track have a velocity of $\vec{v}$, and let the reconstructed track pass through point $\vec{x}_0$ at time $t_0$.  Then, linefit reconstruction solves the  \textit{least-squares} optimization problem:
\begin{equation}\label{equ:least squares}
\min_{t_0, \vec{x}_0, \vec{v}} \sum_{i=1}^N \rho_i(t_0, \vec{x}_0, \vec{v})^2,
\end{equation}
where 
\begin{equation}\label{equ:def rho}
\rho_i(t_0, \vec{x}_0, \vec{v})=  \left\|\vec{v}(t_i - t_0) + \vec{x}_0 - \vec{x}_i \right\|_2.
\end{equation}
Linefit  is an approximation primarily used to generate an initial track or \textit{seed} for a more sophisticated reconstruction.   

The reconstruction algorithm for the sophisticated reconstruction is  \textit{Single-Photo-Electron-Fit (SPE fit)}~\cite{detectorBackground}.  SPE  fit uses the least-squares reconstruction, the event data, and a parameterized probability distribution function of scattering in ice~\cite{detectorBackground} to reconstruct the muon track.  The SPE fit is the primary reconstruction algorithm used in the initial data selection and filtering run
at the detector site, and the fit serves as a seed track to the more complex reconstructions used in off-site data analyses.

\subsection{Algorithm Improvement}

If angular deviations of the initial seed are large ($>$5-10 degrees), the
       simple subsequent reconstruction, SPE, often does not converge to the 
       global minimum, and the efficiency is degraded. This can be resolved 
       by more advanced but time-consuming reconstruction algorithms or by 
       improving the initial seed, as described here.

As indicated in Equation \ref{equ:least squares}, a least-squares fit models the muon as a single point moving in a straight line, and hits are penalized   quadratically in their distance from this line.  Thus, there is an implicit assumption in this model: that all the hits will be near the muon.  This assumption has several pitfalls:
\begin{enumerate}
\item  It doesn't account for the distinct Cherenkov emission profile from the muon.
\item It ignores the scattering effects of the ice medium.  Some of the photons can scatter for over a microsecond, which means that when they are recorded by a PMT, the muon will have traveled over 300 m away.  
\item While the noise reduction steps remove most of the outlier noise, the noise hits that survive can be far from the muon.  Because these outliers are given a quadratic weight, they exert a huge influence over the model.  
\end{enumerate}

The first two pitfalls occur because the model is incomplete and does not accurately model the data, and the third demonstrates that the model is not robust to noise.  The  solution to this is twofold: improve the model and  increase the noise robustness by replacing least squares with robust statistical techniques.  

\subsubsection{Improving the Model}
	While disregarding the Cherenkov profile is inherent to the simplified model 
        chosen for speed reasons, removing hits generated by photons that 
        scattered for a significant length of time will mitigate the effect 
        of ignoring the photon scattering in the ice.   We found that a basic filter could identify these scattered hits, and improve  accuracy by almost a factor of two by removing them from the dataset.  
	
	More formally, for each hit $h_i$, the algorithm looks at all neighboring hits within a neighborhood of $r$, and if there exists a neighboring hit $h_j$ with a time stamp that is $t$ earlier than $h_i$, then $h_i$ is considered a scattered hit, and is not used in the basic reconstruction algorithm.  Optimal values of $r$ and $t$ were found to be 156~m and 778~ns by tuning them on simulated muon data with an $E^{-2}$ power law spectrum.  
	
\subsubsection{Adding Robustness to Noise }
As described in Equation \ref{equ:least squares}, the least squares model gives outliers quadratic weight, whereas we would prefer that outliers had zero weight.  There are robust models in classical statistics designed to  marginalize outliers. We determined that replacing the least-squares model with a Huber fit~\cite{huber} improves the reconstruction accuracy.

More formally, we replace Equation \ref{equ:least squares} with  the optimization problem:
\begin{equation}\label{equ: min Huber penalty}
\min_{t_0, \vec{x}_0, \vec{v}} \sum_{i=1}^N \phi(\rho_i(t_0, \vec{x}_0, \vec{v})),
\end{equation}
where the Huber penalty function $\phi(\rho)$ is defined as 
\begin{equation} 
\phi(\rho) \equiv \left\{
\begin{array}{ll}
 \rho^2 & \textrm{if } \rho < \mu\\
 \mu( 2\rho-\mu) & \textrm{if } \rho \ge \mu
\end{array}\right..
\end{equation} 
Here, $\rho_i(t_0, \vec{x}, \vec{v})$ is defined in Equation \ref{equ:def rho} and $\mu$ is a constant calibrated to the data (on simulated muon events with an $E^{-2}$ power law spectrum, the optimal value of $\mu$ is 153~m).  

The Huber penalty function has two regimes.  In the near-hit regime  ($\rho < \mu$), hits are assumed to be strongly correlated with the muon's track, and  the Huber penalty function behaves like least squares, giving these hits quadratic weight.  In the far-hit regime  $( \rho \ge \mu)$, hits  are given linear weights, as they are more likely to be noise.   

	In addition to its attractive robustness properties, the Huber fit's weight assignment also has the added benefit that it inherently labels points as outliers (those with $\rho \ge \mu $).  Thus, once the Huber fit is computed, we can go one step further and simply remove the labeled outliers from the dataset.  A better fit is then obtained by computing the least-squares fit on the data with the outliers removed.  The entire algorithm has a mean runtime that is approximately six times longer than Linefit's mean runtime.  

\subsection{Results}
The goal is to improve the accuracy of the reconstruction in order to better separate neutrinos from cosmic rays.  Thus, we present three measurements: (1) the accuracy change between linefit and the new algorithm; (2) the accuracy change when SPE is seeded with the new algorithm instead of linefit; and (3) the improvement in separation between neutrinos and cosmic rays.  

To measure the improvement generated by the changes, we use the metric of \textit{median angular resolution} $\theta_{med}$.  The angular resolution of a reconstruction is the arc-distance between the reconstruction and the true track. The dataset is drawn from simulated neutrino data and is designed to be similar to that observed by the detector.  

\begin{table}[tbp]
\centering
\caption{Median angular resolution (degrees) for reconstruction improvements\label{table:reco results}. The first line is the accuracy of the prior least-squares model, and the subsequent lines are the  accuracy measurements from cumulatively adding improvements into the basic reconstruction algorithm.   }
\begin{tabular}
{|l|c|c|} \hline 
Algorithm& $\theta_{med}$  \\ \hline
Linefit Reconstruction (Least-Squares) &9.917 \\ \hline
With Addition of Logical Filter&  5.205 \\ 
With Addition of Huber Regression& 4.672  \\ 
With Addition of Outlier Removal &4.211\\
\hline 
\end{tabular}

\end{table}

We can improve the median angular
resolution of the basic reconstruction by 57.6\%, as  shown in Table \ref{table:reco results}.  Seeding  SPE with the improved basic reconstruction generates an improvement in the angular resolution of
12.9\%.  These improvements in the reconstruction algorithm result in 10\% fewer
atmospheric muons erroneously reconstructed as up-going, and 1\% more
muons correctly reconstructed as up-going.

\section{Coincident Event Improvements}\label{sec:coincident}

In the second study, we look at the problem of determining when more than one muon has entered the detector.  In the most common case, a single muon will pass though the detector and generate an event before exiting.  These events are processed by the pipeline described in Figure \ref{fig:simple reconstruction pipeline}.  However, for roughly 9\% of the events collected by the data collection algorithm, more than one muon will be passing though the detector simultaneously, an occurrence known as a \textit{coincident event}.  

\begin{figure}[t]
\centering
\epsfig{file=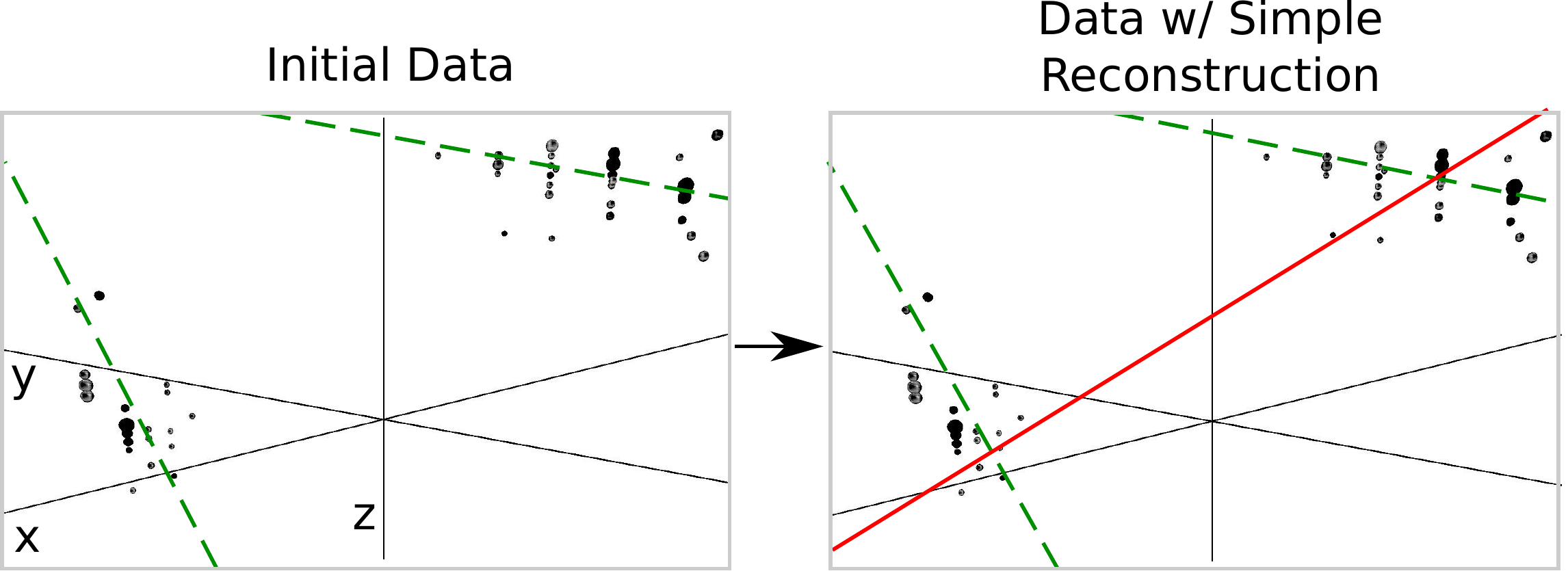, width=12 cm}
\caption{ In this example, an event that is clearly composed of two muons (actual tracks shown as dashed lines) is treated as a single muon, and, thus, the  reconstruction (sold line) is  inaccurate.     \label{fig:bad coincident event reconstruction} }
\end{figure}

\begin{figure*}[t]
\centering
\epsfig{file=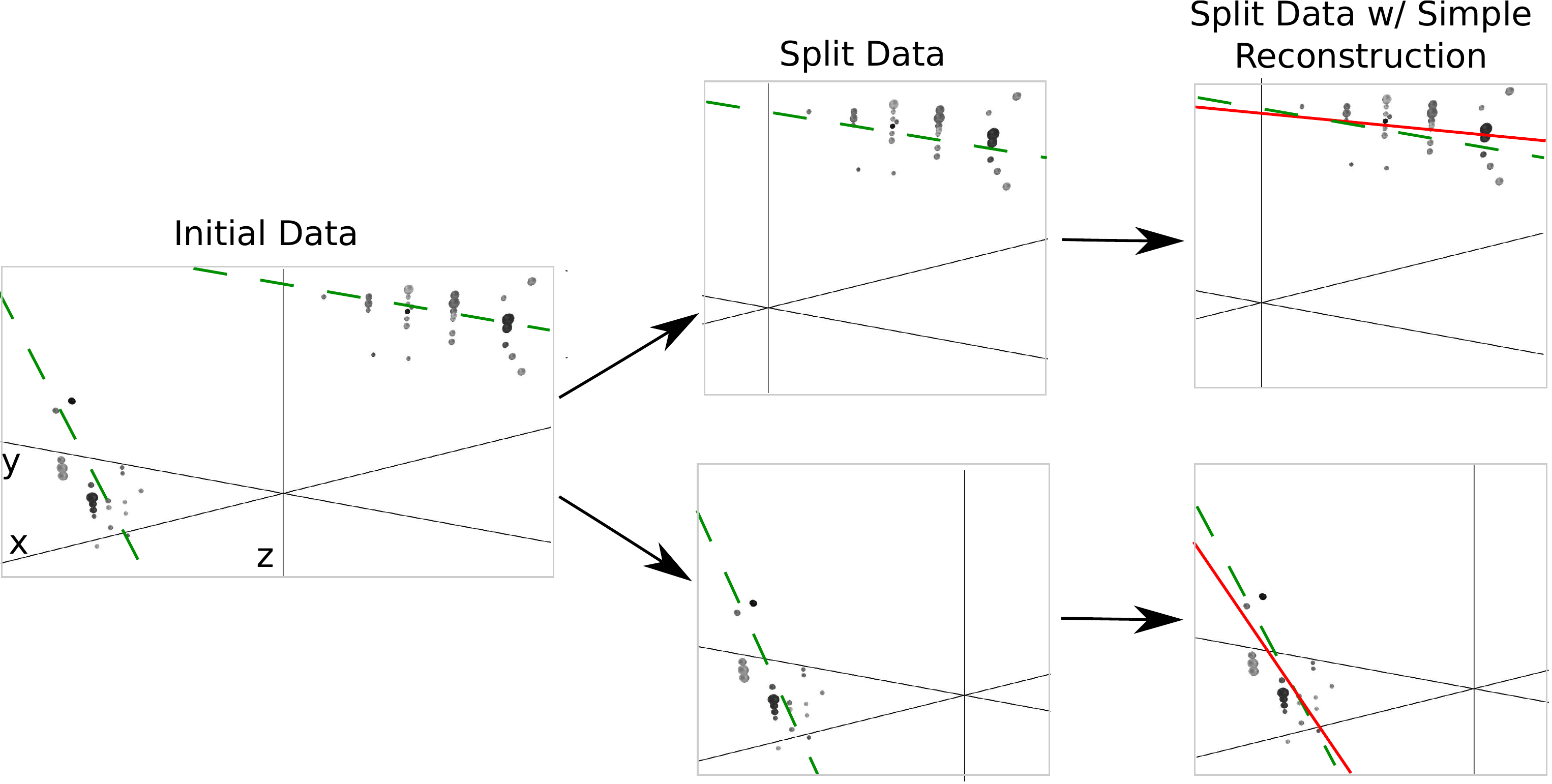, width=12 cm}
\caption{ Ideally, the detector would split coincident events before computing the reconstruction.  Splitting the event results in more accurate reconstructions (reconstructions shown as solid lines, true muon tracks  shown as dashed lines).  Note the difference in the reconstructions compared with Figure \ref{fig:bad coincident event reconstruction}.        \label{fig:split coincident event reconstruction} }
\end{figure*}

One of the primary sources of background noise in IceCube analyses is coincident background muons that have been erroneously reconstructed as neutrinos.  To see why this occurs, consider the coincident event shown in Figure \ref{fig:bad coincident event reconstruction}.  There are two clear groups of hits; however, the reconstruction algorithm treats them as a single group, resulting in an erroneous reconstruction.  In the ideal case, the reconstruction algorithm would identify coincident events and split them, as in Figure \ref{fig:split coincident event reconstruction}. 

The challenge in this example is determining the number of muons in an event.  Our studies show that a simple spatial clustering algorithm can solve this classification problem with less than 2\% error.  

\subsection{Prior IceCube Software}
Coincident events have been a concern in the IceCube analysis~\cite{moreDetectorBackground} for years, and some software has been developed to handle coincident events.  As a baseline of comparison, we use the \textit{TTrigger} software, which  is described in \cite{ttrigger}.  

\subsection{Algorithm Improvement}
  Here, we present a proximal clustering algorithm.  
The intuition in proximal clustering is that points local in space and time are probably from the same muon.  The proximal clustering algorithm iterates through each pair of hits $(i,j)$ and builds an adjacency matrix $\mathbf{A}$ as 
\begin{equation}
\mathbf{A}_{ij} =  \left\{
    \begin{array}{rl}
       1 & \textrm{if } \|\Delta x^2 + \Delta y^2 + \Delta z^2 + (c\Delta t)^2  \|_2 \le \alpha ,\\
      0 & \textrm{otherwise }
    \end{array} \right.
\end{equation}
where $\Delta x, \Delta y, \Delta z$, and  $\Delta t$ are the space and time differences between the pair of hits, and $\alpha$ is tuned to the data (in this application, the optimal value of $\alpha$ is 450~m). The clustering can be recovered by extracting the connected components of the graph defined by $\mathbf{A}$.  A connected component of a graph is a subgraph such, that there exists a path between any two vertices of this subgraph.

\subsubsection{Improving the Model}
When implemented naively, proximal clustering succeeded for the majority of the events, but it failed if there was a gap in the muon track, which can occur when the muon travels through dusty ice layers with short scattering length.  If there is a significantly large gap,  the algorithm erroneously separates the hits into two clusters.

To compensate, an additional heuristic is added, \textit{track connecting}.  After the data segmentation is finished, track connecting determines if separate clusters should be combined.  It computes the mean position and time of each cluster, and connects a hypothetical muon track $T$ between each pair of subspaces.  

It checks if the speed $s$ of the hypothetical track is within 25\% of the speed of light $c$, and it checks that the mean distance between hits and $T$ in both clusters is less than 60 m.  If $T$ passes both checks,  the clusters are combined.  

\subsubsection{Adding Robustness to Noise}
Proximal clustering is susceptible to noise.  Noise hits close to two disjoint tracks will be considered adjacent to both tracks and, thus, can connect the two tracks in the adjacency matrix.  

One heuristic that worked well at mitigating this problem was to not use all of the hits in building the adjacency matrix.  During data collection, some hits are flagged as having a \textit{local coincidence condition}, which indicates that both they and a neighboring PMT reported a hit.  These hits have a high probability of not being noise hits and, thus, exclusively using them to build the adjacency matrix mitigates the problem of erroneously connecting two tracks.  

After the proximal clustering algorithm has extracted the tracks from the adjacency matrix, the hits not used in the construction of the adjacency matrix are simply assigned to the closest reconstructed track.  

\subsection{Results} 
There were two competing goals for coincident event detection algorithms: the algorithm should be conservative enough that events containing single tracks are not erroneously split, and aggressive enough that a useful fraction of coincident events are split correctly.  
Our algorithm is tuned to keep almost all of the single events correctly unsplit, while still correctly splitting  80\% of the coincident events.  

\subsubsection{Measurements}\label{sec: results}
We modified the reconstruction pipeline shown in Figure \ref{fig:simple reconstruction pipeline}, in between the noise cleaning and the basic reconstruction, by adding a step for coincident event detection, as shown in Figure \ref{fig:split coincident event reconstruction}.  This step takes cleaned data and attempts to classify the event as a single-track or multiple-track event.  

We ran each algorithm on two datasets of simulated data.  One dataset comprised single-muon events, and the other dataset comprised multiple-muon events.  In each dataset, we measured the classification error $E$, which is the fraction of events that were misclassified.  To get a global measurement, we  computed the \textit{total error} $E_{tot}$, defined as
\begin{equation}
E_{tot} = w_{\textrm{Single}}E_{\textrm{Single}} + \alpha w_{\textrm{Multiple}}E_{\textrm{Multiple} }.
\end{equation}
For computing $E_{tot}$, we use  $w_{\textrm{Single}} = 0.917$ and $w_{\textrm{Multiple}} =0.083$, which is the frequency in which single-muon and multiple-muon events appear in data simulating the distribution of events that trigger the reconstruction algorithm.   We also include a factor of $\alpha$ in the weighting of the multiple-muon events.  This factor expresses that mischaracterizing a multiple-muon event as a single-muon event degrades the quality of most higher-order analysis more than the reverse mischaracterization.  In our calculations, we use a value of $\alpha = 5$. 

We  present the results for the coincident event problem by measuring how well each algorithm performs at determining the number of subspaces in an event.

There are two natural comparisons for the work: the prior software TTrigger, as well as the trivial algorithm, which always classifies each event as a single-track event.  Clearly, the latter will always get the single-track events correct, and always get the multiple-track events wrong.  We provide a  comparison of these techniques in Table \ref{table:class results}.
\begin{table}[tbp]
\centering
\caption{Error Rates for Classification Algorithms\label{table:class results}}
\begin{tabular}
{|l|c|c|c|} \hline 
Algorithm& $E_{\textrm{Single}}$  \%& $E_{\textrm{Multiple}}$\% & $E_{tot}$ \%    \\ \hline
Trivial &0.0 &100.0&41.5\\
TTrigger &11.5 &31.8&23.7\\
Proximal clustering & 0.2&18.9&8.0\\
\hline 
\end{tabular}
\end{table}
As shown, the new algorithm classifies the number of muons in the detector 66\% better than TTrigger.  

\section{Conclusions}\label{sec:conclusions}
We found that significant improvements can be achieved in the IceCube's online track reconstruction by employing some classical data analysis algorithms. 
Optimizing data filtering and refining the least-square model have led to significant improvements in the accuracy of the reconstruction direction. The new reconstruction software is fast enough to run on-site, and is now included in all IceCube analyses. 

We also looked at the problem of determining the number of muons in the detector. We found that proximal clustering with some basic heuristics could correctly determine whether an event contained a single muon or multiple muons with less than 2\% error, yielding a 66\% improvement over the prior software.


%
\bibliographystyle{elsarticle-num}
\bibliography{cebibliography}  
%
%

\end{document}